\documentclass[10pt,twocolumn]{article}
\usepackage[top=1.8cm, bottom=1.8cm, left=1.8cm, right=1.8cm]{geometry}
\usepackage{helvet}  %
\usepackage{courier}  %

\usepackage{booktabs} %
\usepackage{amsmath,amssymb,amsfonts}
\usepackage{algorithmic}
\usepackage[keeplastbox]{flushend}
\usepackage{graphicx}
\usepackage{textcomp}
\usepackage{xcolor}
\usepackage{multirow,tabularx}
\usepackage[small,bf]{caption}
\usepackage{subcaption}
\usepackage{cleveref}

\newcommand{\descr}[1]{\vspace{0.1cm}\noindent{\bf #1}}

\usepackage{times}

\usepackage[compact]{titlesec}
\titlespacing*{\section}{0pt}{*3}{3pt}
\titlespacing{\subsection}{0pt}{*2}{2pt}
\titlespacing{\subsubsection}{0pt}{*2}{2pt}

\renewenvironment{thebibliography}[1]{
  \begin{oldthebibliography}{#1}
    \setlength{\itemsep}{0.5em}
    \setlength{\parskip}{0.0em}
}
{
  \end{oldthebibliography}
}

\usepackage[hang,flushmargin]{footmisc}

\usepackage[hyphens]{url}  %

\makeatletter
\def\url@leostyle{%
  \@ifundefined{selectfont}{\def\UrlFont{}}%
  {\def\UrlFont{}}%
}
\makeatother
\usepackage[hyphenbreaks]{breakurl}

\definecolor{darkred}{RGB}{153,0,0}
\definecolor{darkblue}{RGB}{0,0,99}
\usepackage[colorlinks=true, linkcolor = darkred,   citecolor = darkred, urlcolor = darkblue]{hyperref}

\renewcommand{\footnoterule}{%
  \kern -3pt
  \hrule width 1in
  \kern 2pt
}

\begin{document}
\sloppy 

\title{\bf LOBO -- Evaluation of Generalization Deficiencies in\\Twitter Bot Classifiers\thanks{Published in the Proceedings of the 2018 Annual Computer Security Applications Conference (ACSAC 2018).}}
\author{Juan Echeverr\'ia$^1$, Emiliano De Cristofaro$^1$, Nicolas Kourtellis$^2$,\\Ilias Leontiadis$^2$, Gianluca Stringhini$^3$, and Shi Zhou$^1$\\[0.5ex]
{\normalsize $^1$University College London $\;\;^2$Telefonica Research $\;\;^3$Boston University}}
\date{}

\maketitle

\begin{abstract}
Botnets in online social networks are increasingly often affecting the regular flow of discussion, attacking regular users and their posts, spamming them with irrelevant or offensive content, and even manipulating the popularity of messages and accounts.
Researchers and cybercriminals are involved in an arms race, and new and updated botnets designed to defeat current detection systems are constantly developed, rendering such detection systems obsolete.

In this paper, we motivate the need for a generalized evaluation in Twitter bot detection and
propose a methodology to evaluate bot classifiers by testing them on {\em unseen} bot classes.
We show that this methodology is empirically robust, using bot classes of varying sizes and characteristics and reaching similar results,
and argue that methods trained and tested on single bot classes or datasets might not able to generalize to new bot classes.
We train one such classifier on over 200,000 data points and show that it achieves over 97\% accuracy.
The data used to train and test this classifier includes some of the largest and most varied collections of bots used in literature.
We then test this theoretically sound classifier using our methodology, highlighting that it does not generalize well to unseen bot classes.
Finally, we discuss the implications of our results, and reasons why some bot classes are easier and faster to detect than others.
\end{abstract}

\section{Introduction}

Automated malicious activity on social networks such as Twitter has been a significant problem for many years now.
Fake accounts controlled by bots are used to perform various types of abuse, e.g., sending spam~\cite{gao2010detecting,grier2010spam}, participating in reputation-manipulation schemes~\cite{boshmaf2011socialbot,de2014paying,ikram2017measuring,stringhini12twitterfollowermarketWOSN}, spreading malware~\cite{egele2013compa}, and phishing \cite{echeverria2017bursty}.
Large quantities of malicious accounts are often created and controlled by single miscreants, forming so-called botnets~\cite{abu2006multifaceted}.
To counter this problem, the research community has developed a number of systems to detect and block bot accounts on social networks.
Such approaches look at either profile characteristics of fake accounts that distinguish them from legitimate ones~\cite{benevenuto2010detecting,stringhini2010detecting}, at differences in the social graph of fake and legitimate accounts~\cite{cai2012latent,danezis2009sybilinfer,liu2015exploiting}, at the way in which they are controlled by their operators~\cite{cao2014uncovering,stringhini2015evilcohort,wang2013you}, or at the content that they post, looking for signs of maliciousness~\cite{lee2012warningbird,nilizadeh2017poised,thomas2011design}.

Despite the large body of research on detecting bots on Twitter, this is still an open problem. 
One reason for this is that bot detection is an inherently adversarial problem, and once a defense mechanism is known, adversaries can modify their modus operandi and avoid detection~\cite{yang2011free}.
Another reason, more fundamental, and often overlooked by the research community is that detection systems  based on machine learning require example datasets of bots to be trained on, and these often contain biases.
For example, if a system was trained on a dataset containing only bot accounts belonging to one botnet, it would learn the idiosyncrasies (e.g., the times at which messages are typically sent or the spam templates used) of that specific botnet and become very accurate in detecting it.
However, when trying to identify bots belonging to other botnets it would perform very poorly, because rather than learning the general characteristics of bots on Twitter, it would overfit on a single family of bots.
Even having multiple families of bots represented in the training set, there is no guarantee that the system will be able to identify new bot types or new botnets as they appear.

In this paper, we set to study this problem in a systematic way. Firstly we collect a dataset that contains more than 20 different bot classes , most of them used in previous bot detection efforts as ground truth \cite{cresci_fame_2015,cresci_paradigm-shift_2017,gilani_bots_2017,echeverria2017bursty,echeverria2017thestar}. Secondly,
we propose a methodology to overcome this issue and produce a generalized bot detection method. This methodology takes into account multiple types of bots, and leverages state-of-art machine learning algorithms for detection of different types of bots.
The training and testing we introduce is done using an effective ``Leave-One-Botnet-Out'' (LOBO) method, which allows the machine learning algorithms to train on data produced by many and diverse bots, and test its accuracy on datasets which include bots with behaviors never seen before by the classifiers. 

In particular, we use this novel methodology on these classes of Twitter bots testing on over 1.5 million bots.
We show that the typical approach of training a model to detect bots using single bot dataset is extremely effective, effortlessly reaching $>$97\% accuracy.
However, the way these datasets are collected prevents them from being representative of all bots in Twitter. We demonstrate that when we mix bot classes equitably in a single dataset, the prediction power of the same classifier drops significantly.
More importantly, we demonstrate that even this bot-detection system that has been trained with a variety of bots is incapable of detecting new bot families that were never observed before.
In fact, some ``target'' botnets completely mislead the classifier resulting to less than 1\% detection accuracy, meaning 99\% of the bots in that class were classified as users.

This methodology provides a proxy for the real world generalization performance of the bot classifier being evaluated.
It further aids in identifying how much each target class is related to the rest of the bot classes without the need of extensive and costly inspection.

Our results provide important insights to the research community including a way to compare bot detection algorithms, beyond their stated accuracy.
We further suggest high generalization performance does not necessarily follow high accuracy.
We finally show the positive insight that even a small portion of certain botnets can be enough to fully identify them when adding it to a common learning algorithm, allowing a classifier to quickly scale, and incrementally consider different and newly revealed bots.
In summary, this paper makes the following contributions:

\begin{itemize}
	\item Shows the need to go beyond common machine learning metrics like accuracy, precision, recall, etc. for Twitter bot detection. As even getting near perfect values for all of them for single bot classes is not necessarily followed by the ability to detect other botnets. 
	\item Addresses that need by providing a framework with which to evaluate expected generalization of a bot detection algorithm by selectively leaving bot classes and behaviours out of the training data.
	\item Collects and combines the biggest and broadest botnet library to date, which it uses to train a Twitter bot classifier.
	\item Provides a Twitter bot classification strategy that reaches accuracy values over 97\% with a small number of commonly used features, and evaluates its performance using the generalization test mentioned earlier.
	\item Introduces a framework to explore the trade-offs between adding more data from a single bot class and diversifying the training data with data from a different bot class. Then analyzes the amount of new samples needed to reach reasonable performance on an target bot class, and discuss on why differences in this metric happen.

\end{itemize}

\section{Related Work}

\descr{Bot Detection.} Early approaches to detect bots on Twitter rely on account characteristics that are typical of fake accounts~\cite{benevenuto2010detecting,stringhini2010detecting}.
Yang et al.~\cite{yang2011free} show that these approaches have a hard time keeping up with the evolution of bots, and that they require constant retraining.
Another line of work looks at the way in which bots connect with other accounts, forming social networks that are very different from the ones built by legitimate accounts~\cite{boshmaf2015integro,cai2012latent,danezis2009sybilinfer}.
However, Liu et al~\cite{liu2015exploiting} show that these techniques can be gamed by adversaries by exploiting the temporal dynamics used for detection.

Other approaches leverage bots' similarity in their operation, such as synchronization in posting messages~\cite{cao2014uncovering}, accesses by a common set of IP addresses~\cite{hooi2016fraudar,stringhini2015evilcohort}, or similar uses of the accounts~\cite{wang2013you}.
Additional work focuses on the content posted by bots.
Thomas et al.~\cite{thomas2011design} analyze the content of the Web pages linked by tweets, learning to identify signs of spam.
Lee et al.~\cite{lee2012warningbird} look for signs of evasion commonly used by cybercriminals, e.g., multiple HTTP redirections.
More recently, Nilizadeh et al.~\cite{nilizadeh2017poised} presented \textsc{poised}, a system that detects malicious messages (e.g., spam) by identifying ones with anomalous spreading patterns across the Twitter graph.

\descr{Fake Accounts.} Thomas et al.~\cite{thomas2011suspended} analyze over a million accounts suspended by Twitter and,
in follow-up work~\cite{thomas13traffickingfraudtwitteraccounts} trafficking of fake Twitter accounts.
Yang et al.~\cite{yang12spammersocialnetwork} look at social relationships between spam accounts on Twitter, while Dave et al.~\cite{dave2012measuring}  measure click-spam in ad networks, and Gao et al.~\cite{gao2010detecting} analyze spam campaigns on Facebook.
Stringhini et al.~\cite{stringhini12twitterfollowermarketWOSN,stringhini13twitterfollower} study the market of Twitter followers and proposes strategies to detect them in the wild.
Finally, there have been a few efforts both in and out of academia to identify single botnets in their entirety. Some of them have obtained large botnets based on geographic anomalies \cite{echeverria2017thestar} and temporal anomalies \cite{echeverria2017bursty}. There have been also analysis on botnets promoting topics or products, including diet pills \cite{narang_satnam_green} and even political candidates \cite{bessi_social_2016}.

Our work takes features from these efforts, but changes one important aspect of them. We explicitly test against unseen classes to evaluate the performance of a classifier.
We create a test for this, which is inspired by cross validation and is thought of as a proxy for generalization of bot classification.
It will be clear that for a bot detection strategy to be deemed as ``generalizable'', the \emph{minimum} standard that should be passed is the test designed in this paper.

\section{Datasets}

Two  datasets were compiled for this paper. First, \emph{a botnet dataset} that contains the aggregated content generated from a variety of bot datasets (some previously used in research as ground truth \cite{cresci_fame_2015,cresci_paradigm-shift_2017,gilani_bots_2017})  and, second, a \emph{real-user dataset}.

\begin{figure*}[ht]
	\includegraphics[width=18cm]{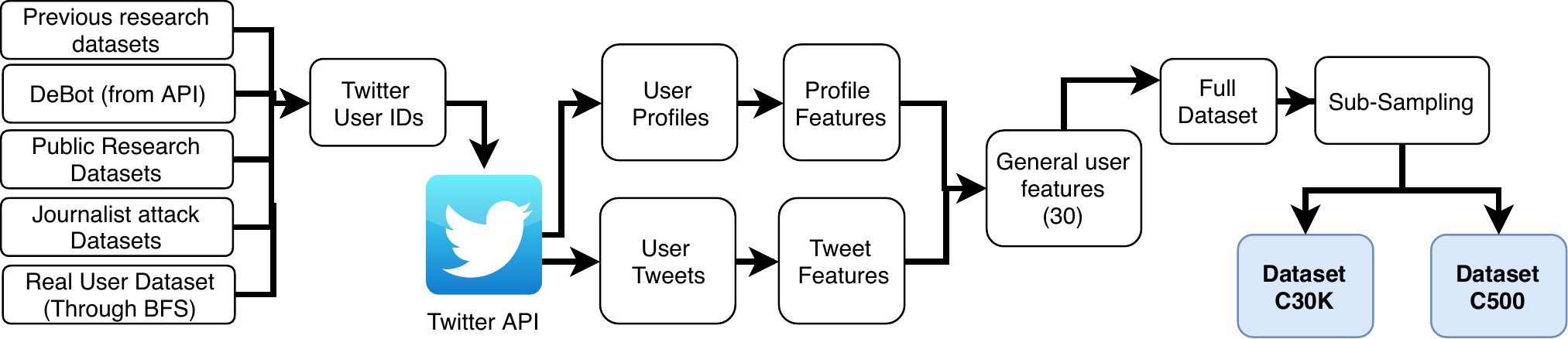}
	\caption{Data collection strategy.}
	\label{Fig. Data Collection Flowchart}
\end{figure*}

Each dataset includes the information available from the user's profile, and all the retrievable tweets at collection time in accordance to Twitter's API limitations.
This means that each account in our dataset contains a maximum of 3,200 tweets authored or retweeted by that account.
The way these datasets were finally constructed is illustrated in Fig.~\ref{Fig. Data Collection Flowchart}.

\subsection{Bot Datasets}
In this work, we study to which extent various bot types have different signatures that can potentially lead to detection failure when they are first discovered (before enough of their samples are identified and included to the training set).
To do this, we build a dataset of 20 different botnet types, each with different purpose and characteristics.
To our knowledge, this is the most extensive and diverse collection of Twitter bots used so far in the literature.

While most of the datasets have high percentage of  bot accounts associated with them, a few might have small amounts of false positives in them. This is a trade-off that needs to be made to avoid inducing bias by filtering these datasets before classification.
Do note that we are only reporting the number of accounts that, at collection time, had not been suspended by Twitter or marked as private. This is because suspension or marking an account as private prevents us from collecting any of its information.

Many of these datasets were obtained from past papers either from authors or directly querying each of the user IDs in them against the Twitter API. Some of them come from an API themselves, and a couple of extra ones are not related to research, but the result of botnet attacks on two journalists and their own listing of the IDs that were involved in those attacks. A summary of these datasets is presented in Table~\ref{tab:Datasets}. Overall, what follows is how we aggregated one of the largest and most varied bot datasets in research.

\descr{The Star Wars Bots.} Dataset \textbf{A} consists of bot accounts that tweet exclusively quotes from Star Wars novels. It was reported by \cite{echeverria2017thestar}. The Star Wars bots all share characteristics like a creation period, id range and small numbers of friends and followers. This dataset consists of over 355,000 accounts.

\descr{The Bursty Bots.}
The Bursty Bots is a botnet created on Twitter with the objective of enticing users into blacklisted sites \cite{echeverria2017bursty,echeverria2018BotnetAttack}. It's strategy was simple by using a mention and a shortened or obfuscated URL. They share some characteristics like having zero friends and followers, and only a few tweets created immediately after account creation, only to remain completely silent afterwards. This dataset \textbf{B} consists of over 500,000 accounts. 

\descr{DeBot.}
Debot is a bot detection service that generates daily reports of bot activity, and stems from the work of \cite{chavoshi_identifying_2016}\cite{chavoshi_debot_2016}.
It comes with an API which we were able to query to obtain over 700,000 accounts that the service detected as bots.
This dataset \textbf{C} is unique in our list, as it is actually the \emph{result} of a detection strategy, and not either ground truth or a single botnet.
The main feature that DeBot detection exploits is warped correlation in the tweet timing of different accounts.

\descr{Fake Followers.}
We explore different fake follower datasets that have been used in various research studies.
Dataset \textbf{D} is used in \cite{cresci_paradigm-shift_2017} and is just described as being fake followers. In contrast, datasets \textbf{Q-T} are described in \cite{cresci_fame_2015} as being purchased fake followers from different fake follower services (\textbf{Q}) fastfollowerz,  (\textbf{S}) intertwitter, and (\textbf{T}) twittertechnology. All these datasets are used as ground truth.

\descr{Traditional spambots.}
Datasets \textbf{H} and \textbf{I} are traditional spam campaigns, pushing links to scam sites. 
Unfortunately the former dataset was unavailable for collection, due to all but one of its accounts being suspended.
Datasets \textbf{K} and \textbf{J} are both groups of accounts spamming job offers.
All these datasets were used in \cite{cresci_paradigm-shift_2017}, while \textbf{H} was also used in \cite{yang12spammersocialnetwork}.

\descr{Social spambots.}
Social spambots are a relatively new breed of bots which are better described in \cite{cresci_paradigm-shift_2017}.
In summary, Social spambots have evolved to accurately mimic the characteristics of real users, making them very difficult to identify.
Dataset \textbf{F} are retweeters of an Italian political candidate.
Dataset \textbf{E} consists of spammers of paid apps for mobile devices.
Finally, dataset \textbf{G} is made of spammers of products on sale at Amazon.com.

\descr{Honeypot bots.}
Dataset \textbf{V} consists of bots collected using honeypot accounts.
A honeypot account is a fake account controlled by a researcher.
The interactions with the account are logged, assuming they can only come from malicious accounts since the honeypot account is fake and generally inactive. 
This dataset was made available through the DARPA twitter bot challenge \cite{subrahmanian_darpa_2016}.
It is used as ground truth in that competition.

\descr{Journalist attack bots.}
In August 2017, journalists Brian Krebs and Ben Nimmo were subject to an attack by twitter bots.
They logged some of the bots and published a dataset \footnote{https://krebsonsecurity.com/tag/twitter-bots/}.
We collected the accounts from these two datasets and added them to our bot datasets as datasets \textbf{W} (the attack on Brian Krebs) and \textbf{X} (the attack on Ben Nimmo).
These datasets are not used as ground truth and, to the best of our knowledge, have not been used in research before.

\descr{Human Annotated Bots.}
Datasets \textbf{L,M,N} and \textbf{O} have been identified by humans as bots, and were used as ground truth in \cite{gilani_bots_2017}.
They were divided by the amount of followers that the bots have.
The bands in which they are divided are 900-1100 followers(\textbf{L}), 90k to 110k followers(\textbf{M}),  900k to 1m followers(\textbf{N}), and  over 9m followers(\textbf{O}). Noticeably, the intermediate groups with different numbers of followers were not available for collection.

\subsection{Aggregated Bot Dataset}
Our aggregated bot dataset is over 1.5 million bots with all their available tweets.
To the best of our knowledge, this is by far the largest bot dataset that has been analyzed in research.
It contains bots from several different sources, including content polluters, fake followers, silent accounts, phishing bots, and political bots (albeit, in a wide array of quantities for each class).

\begin{table}[t!]
\centering
\resizebox{0.99\columnwidth}{!}{
\begin{tabular}{|l|l|r|r|r|}
\hline
\textbf{ID}& \textbf{Name} &\textbf{BTS(\%)}&\textbf{BTS(Avg)}&\textbf{Size}\\\hline
A&Star Wars Bots&---&---&357,000\\\hline
B&Bursty Bots&2.75&0.04&500,000\\\hline
C&DeBot&7.67&0.09&700,000\\\hline
D&Fake followers&96.79&0.90&721\\\hline
E&Social spambots \#1&92.35&0.85&551\\\hline
F&Social spambots \#2&99.37&0.96&3,320\\\hline
G&Social spambots \#3&94.10&0.87&458\\\hline
H&Traditional spambots \#1&98.28&0.93&872\\\hline
I&Traditional spambots \#2&100.00&0.85&1\\\hline
J&Traditional spambots \#3&66.08&0.60&283\\\hline
K&Traditional spambots \#4&97.81&0.90&977\\\hline
L&$\sim$ 1k followers&20.89&0.21&387\\\hline
M&$\sim$ 100K followers&10.90&0.13&534\\\hline
N&$\sim$ 1M followers&1.32&0.02&229\\\hline
O&$\sim$ 10M followers&0.00&0.00&26\\\hline
Q&Fake followers-FSF&100.00&0.96&33\\\hline
S&Fake followers-INT&100.00&0.95&64\\\hline
T&Fake followers-TWT&95.34&0.89&624\\\hline
V&HoneyPot bots (Darpa) &27.69&0.30&2,521\\\hline
W&Attack on Ben Nimmo&59.09&0.54&1,558\\\hline
X&Attack on Brian Krebs&83.05&0.78&728\\\hline
\end{tabular}
}
\caption{Different bot datasets, their identifiers, botometer metrics, and number of accounts collected for each of them}
\label{tab:Datasets}
\vspace{-0.2cm}
\end{table}

\subsection{User Dataset}

To contrast against the bot datasets, we face the  problem of finding a suitable \emph{real}-user dataset of similar size.
While we could just randomly sample Twitter to get an equal amount of users, this methodology might result in including a small amount of bots in our real-user class.
To minimize this issue, we use crawling techniques that attempt to give us an unbiased sample of the general real-user population in Twitter.
Please note that this methodology \emph{does not guarantee} that no bots will be represented in this dataset: it is just a way of minimizing their presence.

We begin with a real user as a seed, and follow his outgoing connections (friends only, not followers).
We manually verify that each of the users in the first level are real users.
We use up to 4 steps and obtain over 1 million English speaking users. 
We assume that a real user is unlikely to follow bots.
A similar approach was used in \cite{echeverria2017thestar}.
While there might be a few bots in this dataset, the vast majority of it must be real users.

\subsection{Botometer Scores}
Botometer \cite{davis_botornot_2016} (previously botornot) is a public API that provides a score based on whether an account is likely to be a bot or a user. It has been used to verify bot accounts in other research \cite{chavoshi_debot_2016}.
For our different bot classes, botometer does not perform well enough. This can be seen in Table~\ref{tab:Datasets}, which shows the average botometer scores for each of the bot classes.
To evaluate whether this tool would be able to predict the dataset, we provide another metric which is the percentage of the queried accounts that receive a botometer score over 0.5 .
Because of rate limiting, we only collected botometer scores for up to 1,000 randomly selected accounts belonging to each of the classes.

We can see that many of these bot classes are overwhelmingly classified as users, for example, only 2.75\% of the Bursty Bots are classified as bots, and less than 10\% of DeBot bots (dataset C) are classified as bots.
Both of them with average botometer scores less than 0.1, indicating that they are ``very likely'' to be users. 

Different bot classes will achieve reasonable and even perfect performance on this bot detection task, but variability between bot classes is very clear.

\section{Methodology - The LOBO test}

\begin{figure}[t]
\centering
\includegraphics[width=6.5cm]{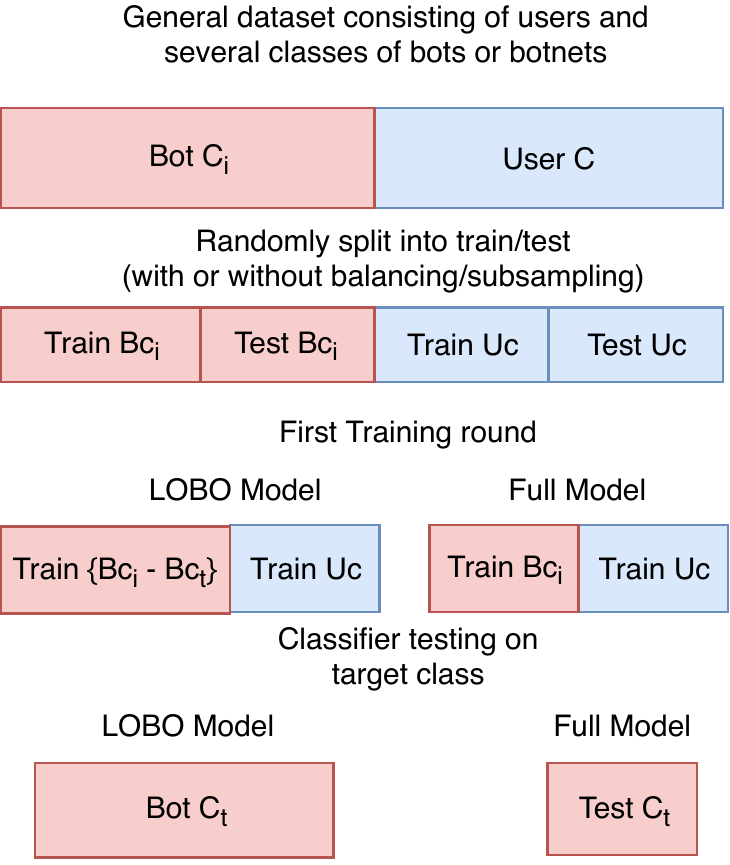}
\caption{Abstract representation of the LOBO test. The classifier gets trained on all bot classes $B_{Ci}$ except the target class $B_{Ct}$, and then tested on the target class to assess how well the classifier generalizes.}
\label{Fig. LOBO flowchart}
\end{figure}

Evaluating bot classifiers faces an important challenge. For obvious reasons, we are unable to evaluate the performance on the bots that we haven't seen. This is a real problem, as bots mutate all the time, and botmasters are actively and creatively trying to get around any form of detection (which sometimes means suspension from the service).

In this paper we propose a new form of testing accuracy for generalization of bot detection strategies. We call it the LOBO test, for \textbf{L}eave \textbf{O}ne \textbf{B}otnet \textbf{O}ut. It derives inspiration from cross validation where a section of the available data is kept out, and used for testing on N number of ``fold.'' Then, the section of the data used for testing changes on each fold. 

The LOBO test was created to assess whether a classifier can detect a bot class, which we will call the \emph{target} class,  by training only with other bot classes, explicitly without any direct knowledge of the target class itself.
It was conceived strictly in the context of a binary classification between bots and users, to specifically address the variety of bot classes that such a classifier would face.

We assume the LOBO test to be a proxy for generalization.
For example, let us take the Bursty bots as the target class. 
We train the classifier with all other datasets except dataset B, then we test the classifier against dataset B.
If the classifier performs well on the Bursty bots when it hasn't actually trained on them, then it has ``generalized'' from the seen bot classes to this target class.
A flowchart of how the test should be applied can be seen in Fig \ref{Fig. LOBO flowchart}.

\descr{Train-Test split.}
We now face the need to compare between a classifier that has been trained with and without the target class.
Addressing this need, we decided to use a 70-30 train-test split instead of cross validation.
Each bot class is randomly and independently sampled so that 70\% of each bot class is in the training set and 30\% in the testing set.
This is to ensure that the smallest classes will still be represented and tested properly.
This strategy allows us to test a single bot class using the 30\% test data for that specific class, which has never been seen by the classifier.

One could argue that comparing accuracy on 30\% of the target class with accuracy tested on 100\% of the target class is misleading. However, this split is not strictly part of the LOBO test, we find it useful to test on at least 30\% of each bot class to provide context to the performance of the classifier on the target class when it has already trained on it. 

\section{Features for Classification}
This section describes the features to be used in our classifier.
Most of these features have been used in research before and are relatively common place.
We placed specific importance on not including graph information.
Twitter imposes strong rate limits on graph information, making it very time consuming to collect the followers of a popular user (e.g., a celebrity or politician) from Twitter's API.
Additionally, any real time implementation of this classifier would need to depend on a few API calls, which does not bode well with the inclusion of graph information.

Table~\ref{Tab. Features} shows all the features that are used in the classifiers analysed in this paper.
Also, the way the datasets at hand were finally prepared for feature extraction is illustrated in Fig.~\ref{Fig. Data Collection Flowchart}.
Next, we define some of the features that might not be straight forward.

\subsection{User Features}

We include several features obtainable directly from a user profile.
Seconds active and days active is the number of seconds and days between account creation and last obtainable tweet.

Their maximum value is 1 month and 3 years respectively, to account for differences in collection dates. While these two features might seem redundant, seconds active is able to detect bots that tweet immediately after creation date and then fall silent, which have been detected in large numbers \cite{echeverria2017bursty,echeverria2017thestar}. "Days active" is better suited to address how long a user has been tweeting. Merging them in a single feature would possibly lose some of the details. Total tweet count is the number of lifetime tweets the account has created or retweeted, and this number comes directly from the profile and is not limited to the 3200 tweets that we get through the API. It includes tweets that have been deleted. We consider these profile features because they can be obtained from a \emph{single} API call to the user profile (which includes the user's last tweet).

\subsection{Tweet Features}

We use the tweets and their text obtained from users to extract useful features.
The most basic ones such as number of tweets and retweets, average tweet and retweet length, etc., are considered, and other more elaborate features are computed and used.

\descr{Hashtags.}
Hashtags are a way of grouping topics within Twitter.
We have created features from the number of hashtags in analyzed tweets and retweets, number of unique hashtags, and the unique hashtag ratio. 
Unique hashtags are included to account for the difference between accounts tweeting many times using a small set of hashtags against people who are involved and tweeting over many different hashtags.

\descr{Mentions.}
Mentions are a way of publicly addressing another user.
They are also commonly abused by spammers to generate engagement with unsuspecting users.
We use the number of mentions in tweets and in retweets as features.
We also include the number of unique mentions as a feature.

\descr{Edit Distance.}
To account for tweets that are equal or with small variations, we use the edit distance between tweets and retweets of a user.
The edit distance (or Levenshtein distance) between two strings is the minimum number of one-character edits to turn one string into the other.
Because of processing time, we only evaluate this feature for the last 200 tweets and the last 200 retweets of each user; each of these tweets is compared to the rest, and then the mean of the distances is computed.

\descr{Geolocation.}
Depending on user preference, each tweet can have geolocation embedded, consisting of latitude and longitude.
We use the number of tweets that are geolocated, as well as the percentage of the analyzed tweets that have this information, as features.

\descr{Tweet Sources.}
When apps are used to publish tweets through Twitter's API, an app publisher needs to define the "source" of the tweet. We use the number of \emph{unique} sources used to publish the tweets as a feature.
While some older botnets rely on using a single source to publish all of their tweets \cite{echeverria2017bursty,echeverria2017thestar}, other botnets may use as many sources as possible to confuse detection efforts.
Regardless of the assumption, we calculate this feature for each of the users in our dataset.

\descr{Favorites.}
Marking a tweet as a favorite or ``liking'' it, is an action a user can take to endorse a specific tweet.
We use the number of tweets a user has liked as a feature.
However, we also include how many of a user's tweets have been marked as favorites by other users, and the ratio between liked tweets and analyzed tweets (or favorites per tweet).
This summarizes both directions of the endorsement: how much the user being analyzed endorses other users, and how much other users endorse the user at hand.

\begin{table}[t]
\resizebox{0.99\columnwidth}{!}{%
\centering
\begin{tabular}{|l|l|l|}
\hline
\multicolumn{3}{|c|}{ \textbf{User Features}} \\\hline\hline
User ID& \# Followers & \# Friends\\\hline
Friend To Follower Ratio & \# User favorites & username length\\\hline
Seconds active& Days active & Total tweet count\\\hline
Profile description length &&\\\hline\hline
\multicolumn{3}{|c|}{\textbf{Tweet Features}} \\\hline\hline
\# Geolocated tweets &  \% of Geolocated Tweets & \# Mentions (Rtw) \\\hline
Avg. Edit distance (Rtw)& \# Hashtags & \# Mentions (Twt)\\\hline
Avg. Edit distance (Twt)& \# URLs & URLs (per tweet)\\\hline
\# Tweets analyzed & \# Favorites & Favorites (per tweet)\\\hline
\# APIs used & \# Unique hashtags & Unique hashtag ratio\\\hline
\# Unique mentions & Avg. tweet length & Avg. retweet length\\\hline
\# Retweets analyzed&&\\\hline
\end{tabular}
}
\caption{Features employed for classification.}
\label{Tab. Features}
\vspace{-0.2cm}
\end{table}

\section{Experiments}

In this section, we describe the experimental setups used, machine learning classifiers applied and results extracted using the LOBO test under different scenarios.

\subsection{Subsampling}

\descr{Dataset with class size $\leq$ 30k}
In real life, botnets will come in varying sizes.
Furthermore, the amount of data that will be available for training each bot class will vary even more. As an easy example, the bot classes analyzed in this paper range from tens of accounts to hundreds of thousands. 
To provide an opportunity for our smallest bot classes, in this experimental setup we limit the numbers of the three larger bot classes.
With this in mind, we include only 30,000 randomly sampled bot accounts from each of these datasets (\textbf{A, B}, and \textbf{C}).
However, we include all of the bots in datasets \textbf{D-X} for a total of $\sim$ 105,000 bots.
The reasoning behind this is to not allow our three large datasets to exceed 100 times the largest of our other, smaller, datasets.

To contrast these bots against users, we randomly sample our real user dataset to only include 105,000 accounts.
The aggregation of these 105k users and 105k bots will be referred to as the dataset with class size $\leq$ 30k, \textbf{C30K} for short.

\descr{Dataset with class size = 500}
Dataset C30K is still quite imbalanced, having classes with 30,000 bots and classes with 26 bots.
To measure the effect of bot classes without being biased by their size, we create another bot dataset. This balanced sub-sampled bot dataset contains 500 random instances from each of the bot classes that have over 500 accounts in them.
This means a few of the bot datasets have been excluded, but choosing 500 as the size still allows us to have 14 bot datasets to evaluate on.
This bot dataset is made of 7,000 bot instances. 

To contrast against this bot dataset, we add an equivalent number of users from our user dataset.
The aggregation of both of these datasets will be referred to as the dataset with class size = 500, or \textbf{C500}.
Please note that this dataset is created on the fly every time it is needed.
Fig.~\ref{Fig. Data Collection Flowchart} shows the data collection flow from the bot class identification to our C30K and C500 datasets.

\subsection{General Classifiers}
We utilize our user dataset against the bot training data.
All the features presented in previous Section have been calculated for every user, making each user representation a 30-dimension vector.
We use some of the most common classifiers (mostly based on trees).
The classifiers to test are Gradient Boosted Trees (using Xgboost and LightGBM), Random Forests, Decision Trees, and AdaBoost.

All these algorithms are deliberately trained using their most standard and naive python implementation. Naturally, the performance evaluation is done purely on the test data which was not ``seen'' during training.

\begin{table}[t]
	\centering
	\small
	\resizebox{0.75\columnwidth}{!}{%
	\begin{tabular}{|l|r|r|r|r|}
		\hline
		\textbf{}  & \multicolumn{2}{c|}{\textbf{C30K}}  & \multicolumn{2}{c|}{\textbf{C500}}  \\ \hline
		\multicolumn{1}{|c|}{\textbf{Algorithm}} & \multicolumn{1}{c|}{\textbf{Acc.}} & \multicolumn{1}{c|}{\textbf{AUC}} & \multicolumn{1}{c|}{\textbf{Acc.}} & \multicolumn{1}{c|}{\textbf{AUC}} \\ \hline
LGBM          & 97.84\% & 0.98 & 93.93\% & 0.94 \\ \hline
XGBC          & 95.91\% & 0.96 & 91.24\% & 0.91 \\ \hline
Random Forest & 97.02\% & 0.97 & 91.54\% & 0.92 \\ \hline
DecissionTree & 95.99\% & 0.96 & 86.93\% & 0.87 \\ \hline
AdaBoost      & 94.29\% & 0.94 & 88.88\% & 0.89 \\ \hline
	\end{tabular}
	}
	\caption{Classifier performance on C30K and C500 datasets.}
	\label{Tab. General Classifier Accuracy}
\end{table}

\begin{table}[t]
\centering
\resizebox{0.9\columnwidth}{!}{%
\small
\begin{tabular}{|l|r|r|r|r|}
\hline
\multicolumn{1}{|c|}{\textbf{}} & \multicolumn{1}{c|}{\textbf{\begin{tabular}[c]{@{}c@{}}True \\ Negatives\end{tabular}}} & \multicolumn{1}{c|}{\textbf{\begin{tabular}[c]{@{}c@{}}False \\ Positives\end{tabular}}} & \multicolumn{1}{c|}{\textbf{\begin{tabular}[c]{@{}c@{}}False \\ Negatives\end{tabular}}} & \multicolumn{1}{c|}{\textbf{\begin{tabular}[c]{@{}c@{}}True \\ Positives\end{tabular}}} \\ \hline
LGBM          & 31161 & 344  & 1010 & 30152 \\ \hline
XGBC          & 30767 & 738  & 1824 & 29338 \\ \hline
Random Forest & 31073 & 432  & 1438 & 29724 \\ \hline
DecissionTree & 30241 & 1264 & 1250 & 29912 \\ \hline
AdaBoost      & 30002 & 1503 & 2078 & 29084 \\ \hline
\end{tabular}
}
\caption{Confusion Matrices for different classifiers trained on dataset C30K.}
\label{Tab. Confusion Matrix for C30K}
	\vspace{-0.2cm}
\end{table}

\descr{Performance Evaluation - C30K}
Table~\ref{Tab. General Classifier Accuracy} shows the results of a binary classification attempt using the dataset C30K.
All of the algorithms show clear signs of an easy separation task, with accuracies over 95\% in most cases.
This level of accuracy in bot classification is not unheard of: it has been claimed before several times (e.g.,~\cite{cresci_fame_2015,subrahmanian_darpa_2016,lee_seven_2011}).

To further reiterate that this is not a fluke, we also check the area under the ROC curve and the confusion matrix for some of the results generated (Table~\ref{Tab. Confusion Matrix for C30K}).
As can be seen, almost all bots are classified as bots, and almost all users are classified as users, for all the algorithms tested (remember that the testing set was 30\% of the C30K dataset, i.e., $\sim$63k instances total).
This result was repeated several times just for consistency, all with random 70-30 training-test splits and showed little variation. One could argue that our LGBM classifier is comparable to the state of the art in bot detection,  having been trained with over 200,000 data points spanning a wide variety of bot classes, achieving accuracy of over 97\%.

\descr{Performance Evaluation - C500}
We need to know the performance on a dataset where the bots have the same numbers, since we cannot always count on having the benefit of large bot data corpuses like DeBot, Star Wars bots or Bursty bots. 

We evaluate performance on dataset C500 with the same strategy, using the same standard and naive versions of several popular classification algorithms.
The results, while still encouraging, show clear deterioration in accuracy.
In Table~\ref{Tab. General Classifier Accuracy}, we see more than 5\% loss in accuracy for the best performing algorithm, and a steeper 8\% loss for decision trees.

\subsection{LOBO Test I - C30K}

We run a LOBO test on our C30K bot dataset with class size $\leq$ 30k. It follows the steps in Fig. \ref{Fig. LOBO flowchart}. The results are summarized in Tab. \ref{Tab. LOBO Full}, where:

\begin{itemize}
  \item \textbf{Target Class} is the dataset ($Bc_t$) or bot class that is being target of training
  \item \textbf{Full Model Accuracy} is the accuracy of the binary model trained on the training set of \emph{all} classes ($Train \; Bc_i + Train \; Uc$). It is calculated on the test subset of the target class ($Test \; C_t$). It provides the \emph{expected} performance of the general (full) model on the target class, this is useful for context.

  \item \textbf{LOBO Model Accuracy} is the accuracy of the model trained on the dataset that excludes the target class ($Train \; \{Bc_i -  Bc_t\} \; + Train \; Uc$) when tested on the full target class ($Bc_t$) . This measure can be tested against the complete target class because none of it has been used for training the model\footnote{Because we are testing \emph{only} on the bot class, accuracy and recall are the same because false positives and true negatives are zero}.

  \item \textbf{Acc. gain} This is just \textbf{LOBO Model Accuracy} subtracted from \textbf{Full Model Accuracy}. It represents how much a model's performance improves when trained on a the target class, as compared to its performance without training on the target class. It is a subtraction because differences will be substantial, so a ratio would have been misleading. 
\end{itemize}

The results speak for themselves. The average expected accuracy on a target bot class that the classifier has not been trained on is 54.88\%. This is almost as bad as random (although we are deliberately excluding the user class from the testing). As would be expected, there are some exceptions that perform well like the Bursty bots (\textbf{B}). However, even before excluding the target class from the training data some of the classes performed as poorly as 19\%. 

It is noticeable that some of these classes are actually "losing" accuracy when being included in the test set, this is most likely due to the large difference in size between the test set for the LOBO model ($Bc_t$) and the test set for the Full model ($Test C_t$). We further note that the average accuracy for all classes is well below the 97\% achieved originally, which only means that we are performing better on the large classes than the smaller ones.

\begin{table}[]
	\centering
	\resizebox{0.75\columnwidth}{!}{%
\small
	\begin{tabular}{|c|r|r|r|}
		\hline
		\multicolumn{1}{|c|}{\textbf{\begin{tabular}[c]{@{}c@{}}Target\\ Class\end{tabular}}} & \multicolumn{1}{c|}{\textbf{\begin{tabular}[c]{@{}c@{}}Full Model\\ Accuracy\end{tabular}}} & \multicolumn{1}{c|}{\textbf{\begin{tabular}[c]{@{}c@{}}LOBO Model\\ Accuracy\end{tabular}}} & \multicolumn{1}{c|}{\textbf{\begin{tabular}[c]{@{}c@{}}Accuracy\\ Gain\end{tabular}}} \\ \hline
\textbf{A} & 97.41\% & 100.00\% & 60.34\% \\ \hline
\textbf{B} & 97.43\% & 99.98\%  & 97.30\% \\ \hline
\textbf{C} & 98.41\% & 96.19\%  & 76.85\% \\ \hline
\textbf{D} & 97.91\% & 75.11\%  & 68.79\% \\ \hline
\textbf{E} & 97.83\% & 92.21\%  & 23.23\% \\ \hline
\textbf{F} & 97.82\% & 98.30\%  & 0.36\%  \\ \hline
\textbf{G} & 97.84\% & 87.59\%  & 11.79\% \\ \hline
\textbf{H} & 97.97\% & 73.47\%  & 0.69\%  \\ \hline
\textbf{J} & 97.79\% & 90.67\%  & 93.29\% \\ \hline
\textbf{K} & 97.71\% & 97.92\%  & 2.35\%  \\ \hline
\textbf{L} & 97.76\% & 87.39\%  & 89.15\% \\ \hline
\textbf{M} & 97.81\% & 80.62\%  & 76.59\% \\ \hline
\textbf{N} & 97.75\% & 90.54\%  & 71.62\% \\ \hline
\textbf{O} & 97.82\% & 100.00\% & 92.31\% \\ \hline
\textbf{Q} & 97.76\% & 80.00\%  & 69.70\% \\ \hline
\textbf{S} & 97.74\% & 72.22\%  & 73.44\% \\ \hline
\textbf{T} & 97.80\% & 73.45\%  & 69.87\% \\ \hline
\textbf{V} & 98.24\% & 72.80\%  & 47.16\% \\ \hline
\textbf{W} & 97.91\% & 71.67\%  & 64.83\% \\ \hline
\textbf{X} & 97.76\% & 85.71\%  & 79.12\% \\ \hline
	\end{tabular}
	}
		\caption{LOBO test on dataset C30K}
	\label{Tab. LOBO Full}
	\vspace{-0.2cm}
\end{table}

\subsection{LOBO Test II - C500}

One could make the argument that the differences in the performance of these classifiers is due to their large imbalance between their classes. We use dataset C500 to test if this is true.

For this test, there is an expected level of variability, as selecting just 500 instances of the large bot classes leaves large percentages of them outside of the training set.
In the case of the Bursty Bots (\textbf{B}) or DeBot bots (\textbf{C}), over 99.9\% of the class is kept out of the training.
To mitigate this effect, we do this measurement 100 times for each of the target dataset.
Each of these times we:

\begin{itemize}
\item Randomly generate the C500 dataset.
\item Randomly split the resulting dataset 70\% for training and 30\% for testing \footnote{In contrast to LOBO test I, now every bot class gets a similar number of instances for training and testing}.
\item Train a classifier on the training set that has just been created (this is referred to as the \textbf{Full Model})
\item Test the Full model on the 30\% testing set of the class that will be evaluated as target. This gives the \textbf{Full Model Accuracy}. 
\item Remove all instances of the target bot class from the training set.
\item Train a new ``LOBO model'' on the new training set that lacks the target class.
\item Test the accuracy of the LOBO model on the full target class (which was recently removed from the training set), and obtain \textbf{LOBO Model Accuracy}
\end{itemize}

All of these steps are performed 100 times for each of the bot classes.
What results is the ability to evaluate how a model trained on balanced bot classes can be expected to perform against a target bot class which is previously unseen  by this model.
Furthermore, it allows performance comparison for when this model has seen just 500 of the target class against not seeing any, with some surprising results.
Table~\ref{Tab. LOBO Sampled} shows average results per bot class.
In this table we added a new measure for context: 1-Class Model Acc. This provides the accuracy when the model is trained and tested on a single bot class (dividing the data in the same 70/30 split).

Another interesting fact is that in this test the accuracy on unseen classes is almost the same as shown in \ref{Tab. LOBO Full} but the average accuracy of the full model is not. In this test, the per class average accuracy for the full model is very close to the expected 92.1\% shown in Tab. \ref{Tab. General Classifier Accuracy}. It is likely due to the balancing of bot classes.

\begin{table}[t]
\centering
\resizebox{0.92\columnwidth}{!}{%
\begin{tabular}{|c|r|r|r|r|}
	\hline
	\textbf{\begin{tabular}[c]{@{}c@{}}Target\\ Class\end{tabular}} & \multicolumn{1}{c|}{\textbf{\begin{tabular}[c]{@{}c@{}}1 - Class\\ Model Acc.\end{tabular}}} & \multicolumn{1}{c|}{\textbf{\begin{tabular}[c]{@{}c@{}}Full Model\\ Accuracy\end{tabular}}} & \multicolumn{1}{c|}{\textbf{\begin{tabular}[c]{@{}c@{}}LOBO Model\\ Accuracy\end{tabular}}} & \multicolumn{1}{c|}{\textbf{\begin{tabular}[c]{@{}c@{}}Accuracy\\ Gain\end{tabular}}} \\ \hline
		\textbf{A}   & 99.80\%      & 93.90\%   & 62.01\%   & 31.89\%       \\ \hline
		\textbf{B}   & 99.74\%      & 93.92\%   & 98.14\%   & -4.22\%       \\ \hline
		\textbf{C}   & 96.42\%      & 94.22\%   & 84.81\%   & 9.41\%   \\ \hline
		\textbf{D}   & 92.37\%      & 94.17\%   & 86.65\%   & 7.52\%   \\ \hline
		\textbf{E}   & 97.42\%      & 94.11\%   & 49.19\%   & 44.92\%       \\ \hline
		\textbf{F}   & 99.47\%      & 94.02\%   & 0.71\%    & 93.31\%       \\ \hline
		\textbf{H}   & 94.56\%      & 94.92\%   & 2.19\%    & 92.73\%       \\ \hline
		\textbf{K}   & 99.70\%      & 94.05\%   & 1.97\%    & 92.08\%       \\ \hline
		\textbf{M}   & 98.43\%      & 94.25\%   & 66.02\%   & 28.23\%       \\ \hline
		\textbf{T}   & 92.79\%      & 94.25\%   & 88.79\%   & 5.46\%   \\ \hline
		\textbf{U}   & 91.14\%      & 95.16\%   & 39.44\%   & 55.73\%       \\ \hline
		\textbf{V}   & 92.22\%      & 94.57\%   & 77.86\%   & 16.71\%       \\ \hline
		\textbf{W}   & 94.53\%      & 94.21\%   & 89.82\%   & 4.39\%   \\ \hline
		\textbf{Avg.}     & \textbf{96.05\%}       & \textbf{94.29\%}    & \textbf{57.51\%}    & \textbf{36.78\%}   \\ \hline
	\end{tabular}
	}
	\caption{Lobo test on dataset C500}
	\label{Tab. LOBO Sampled}
\vspace{-0.2cm}
\end{table}

\section{Beyond the LOBO test}
\subsection{Relatively Stable Results}

To evaluate whether these results are stable or not, we take the standard deviation of the LOBO model accuracy from LOBO test II. This was made on dataset C500 and there are 14 classes to analyze. Almost all of these target classes show low standard deviation of less than 4\%, meaning that regardless of the way the dataset is sampled and split, the accuracy on each target (unseen) class remains stable. This suggests that the LOBO test will provide consistent results overall. The only two exceptions with a standard deviation above 4\% are the Star Wars bots at 24\% standard deviation, and the Social Spambots \# 1 at 13\% .

\subsection{Learning Rate}

To further analyse the gap of accuracy between the Full Model and the LOBO model, here we measure how fast the LOBO model can improve its performance by moving a few of the target bots, from the test data to the training data. The learning rate is measured on a single sampled dataset from LOBO Test II. For example, take the Bursty bots as the target class. Initially, none of the 500 Bursty bots are included in the training data, and all the 500 form the whole of the test data. 

Consider X as the step size. At the first step, we randomly choose X Bursty bots and remove them from the test data, and then add them to the training data (in addition to the training data of the other bot classes). We train the classifier and record the prediction results. In the second step, again, we randomly choose X Bursty bots from the test data, and move them to the training data. And so on. The test finishes at the 9th step when both the test data and the training data contain about 250 Bursty bots. The step sizes are fixed at 0, 2, 4, 8, 16, 32, 64, 128, and 256 (basically it is a $2^x$ scale but we traded the first step for zero bots which matches LOBO test II).Note that, this being a \emph{single} C500 dataset, the difference in accuracy between the 0 bot case and LOBO test II is expected. We also use the full 500 instances of each of the classes except the target, instead of limiting to 70\%.

We repeat the above process 50 iterations, and then calculate the average prediction accuracy at each step for the target class. Repetition is needed because each of the times different bots from the target class are being sent into the training set, and it affects the overall accuracy differently. Finally, we run the learning rate test for each bot class in Table~\ref{Tab. LOBO Sampled} as a target class.
Detailed results are shown in Table~\ref{Tab. Learning Rate}.

\begin{figure}[t]
\centering
\includegraphics[width=8.2cm]{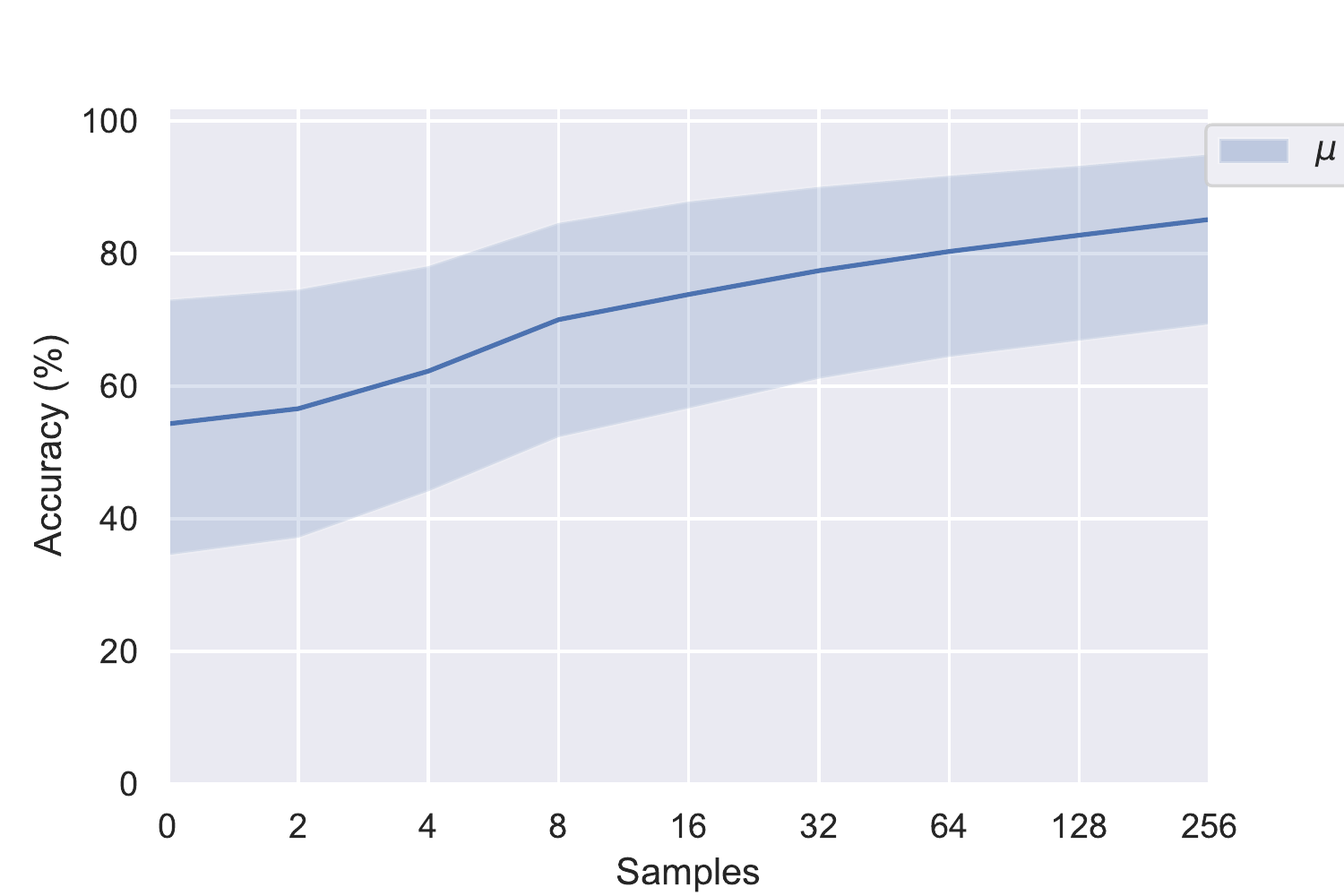}
\vspace{-0.2cm}
\caption{Mean (Blue) and error range (blue shade 95\% confidence) for the classifier accuracy on target classes according to the number of samples seen from the target class}
\label{Fig. Average Performance per Sample (General)}
\vspace{-0.3cm}
\end{figure}

\begin{figure}[t]
\centering
\includegraphics[width=8.2cm]{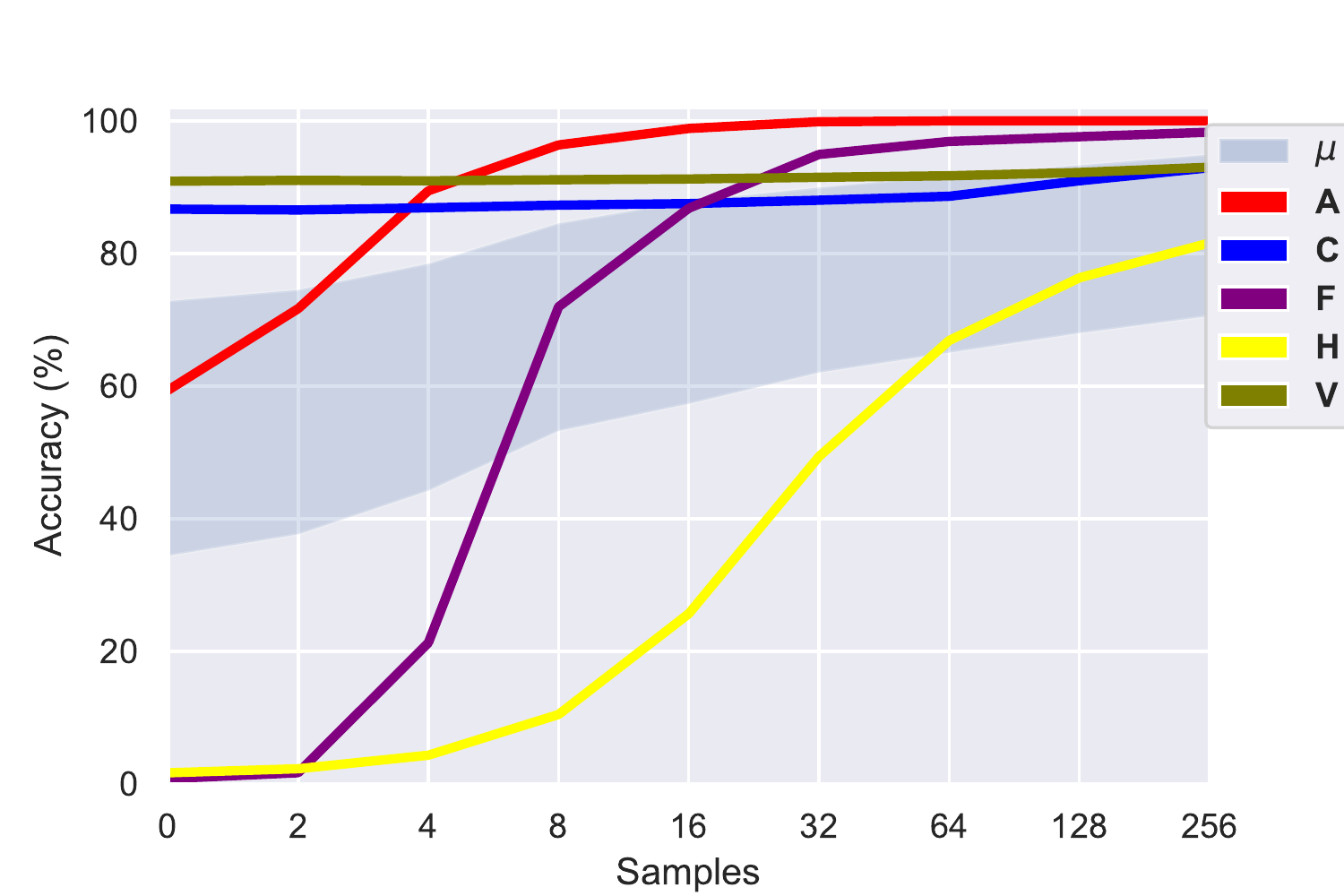}
\vspace{-0.2cm}
\caption{Error range for mean (blue shade) and classifier accuracy against number of samples seen (for specific classes)}
\label{Fig. Average Performance per Sample (Specific)}
\end{figure}

\begin{figure*}[t]
	\centering
\begin{subfigure}[b]{0.8\columnwidth}
	\centering
			\includegraphics[width=8cm]{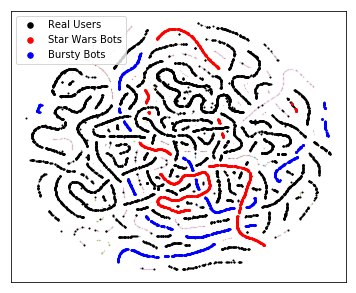}
	\vspace{-0.65cm}
		\caption{Star Wars Bots and Bursty Bots}
		\label{Fig. TSNE SWB}
	\end{subfigure}
	~~~~~~~~~~
\begin{subfigure}[b]{0.8\columnwidth}
	\centering
		\includegraphics[width=8cm]{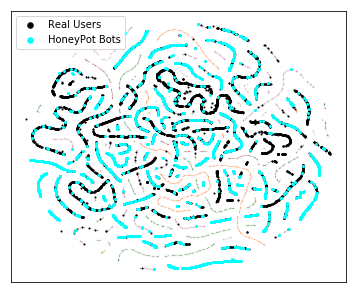}
	\vspace{-0.65cm}
			\caption{HoneyPot Bots}
		\label{Fig. TSNE Bad Performers}
	\end{subfigure}
\caption{T-SNE plot of all the bot classes against real users.}
\end{figure*}

Figure~\ref{Fig. Average Performance per Sample (General)} shows the average accuracy after X samples for all bot classes that have been tested, with the shaded area representing the 95\% confidence interval. The overall trend would suggest that the classifier has learned to identify most target classes of bots after a few examples.
In contrast, Figure~\ref{Fig. Average Performance per Sample (Specific)} shows that the performance for different classes varies significantly.
It contains the same shaded area as Figure~\ref{Fig. Average Performance per Sample (General)} to show the stark differences between the average and the widely varying performance of each target class.

We can further notice bot class \textbf{V} (caught by honeypots) making no reasonable improvement regardless of how many instances of it has been shown to the classifier.
Finally, we can see some classes that increase dramatically from 50 to 99\% accuracy (bot class \textbf{A}) and from 0-95\% accuracy (bot class \textbf{F}), after only 16 and 32 instances, respectively.
This implies learning to identify the whole class while training only on 6\% of it.

Notably, this section has shown the learning speeds for some target classes is much higher than others. Improvements of over 20 accuracy percent points for the addition of 2 single instances are seen in more than one class. Finally, different bot classes show various degrees of improvement. Some of the target classes, worryingly, show almost \emph{no improvement} at all.

\begin{table}[t]
\centering
\resizebox{0.99\columnwidth}{!}{%
\begin{tabular}{|c|r|r|r|r|r|r|r|r|r|}
\hline
\multirow{2}{*}{\textbf{\begin{tabular}[c]{@{}l@{}}Tgt.\\ Class\end{tabular}}} & \multicolumn{9}{c|}{\textbf{Number (X) of samples of target class in training data}}  \\ \cline{2-10} 
 & \multicolumn{1}{c|}{\textbf{0}} & \multicolumn{1}{c|}{\textbf{2}} & \multicolumn{1}{c|}{\textbf{4}} & \multicolumn{1}{c|}{\textbf{8}} & \multicolumn{1}{c|}{\textbf{16}} & \multicolumn{1}{c|}{\textbf{32}} & \multicolumn{1}{c|}{\textbf{64}} & \multicolumn{1}{c|}{\textbf{128}} & \multicolumn{1}{c|}{\textbf{256}} \\ \hline
\textbf{A} & 59.4 & 71.7 & 89.4 & 96.4 & 98.8 & 99.9 & 100  & 100  & 100  \\ \hline
\textbf{B} & 97.4 & 97.4 & 97.7 & 97.4 & 98.1 & 98.6 & 99.0 & 99.6 & 99.8 \\ \hline
\textbf{C} & 86.7 & 86.5 & 86.9 & 87.3 & 87.5 & 88.0 & 88.6 & 90.9 & 92.9 \\ \hline
\textbf{D} & 86.7 & 86.8 & 86.7 & 86.8 & 87.0 & 87.6 & 87.8 & 89.0 & 90.6 \\ \hline
\textbf{E} & 64.9 & 68.6 & 72.2 & 79.8 & 84.9 & 89.4 & 92.8 & 94.7 & 95.9 \\ \hline
\textbf{F} & 0.8  & 1.7  & 21.3 & 72.0 & 86.8 & 94.9 & 96.9 & 97.6 & 98.3 \\ \hline
\textbf{H} & 1.7  & 2.3  & 4.3  & 10.5 & 25.7 & 49.4 & 66.9 & 76.3 & 81.6 \\ \hline
\textbf{K} & 4.0  & 17.8 & 51.9 & 85.3 & 95.2 & 96.6 & 98.5 & 99.2 & 99.5 \\ \hline
\textbf{M} & 64.2 & 64.8 & 66.3 & 67.3 & 70.1 & 74.2 & 80.8 & 87.5 & 95.3 \\ \hline
\textbf{T} & 89.9 & 89.9 & 89.6 & 89.6 & 89.5 & 89.7 & 90.1 & 90.2 & 91.0 \\ \hline
\textbf{U} & 33.7 & 34.0 & 34.6 & 36.9 & 38.5 & 43.2 & 50.0 & 59.1 & 69.6 \\ \hline
\textbf{V} & 80.0 & 79.7 & 79.7 & 79.9 & 80.0 & 80.7 & 81.0 & 82.2 & 84.3 \\ \hline
\textbf{W} & 90.9 & 91.0 & 90.9 & 91.1 & 91.2 & 91.5 & 91.7 & 92.2 & 93.0 \\ \hline
\end{tabular}
}
\caption{Classifier accuracy (\%) - trained C500 excluding all but X samples of target class.}
\label{Tab. Learning Rate}
	\vspace{-0.2cm}
\end{table}

\subsection{TSNE plot}

In an effort to further understand why some of these bot classes seem easier to predict than others, we've created a t-distributed stochastic neighbor embedding (TSNE) plot \cite{maaten_visualizing_2008} with the dataset with class size $\leq$ 30k.
This is a dimensionality reduction algorithm that is also helpful to visualize high dimension datasets in two dimension plots.
In Figure~\ref{Fig. TSNE SWB} it is clearly shown that there are clusters readily formed by different bot classes, where each one is plotted in a different colour, and users are always in black.

Figure~\ref{Fig. TSNE SWB} emphasizes the Star Wars bots and the Bursty bots, which show clear cut groups and clusters that are rarely mixed with the real users.

In contrast, Figure~\ref{Fig. TSNE Bad Performers} tells a different tale. We can see the honeypot bots (dataset \textbf{B}) mostly sharing the same "strands" with the real users. These bots are the ones the LOBO test showed to be difficult to classify, so it is no surprise that they look similar to our user dataset.

\section{Discussion}

\subsection{Accuracy and Generalization}

The average accuracy on target classes was very similar in LOBO tests I and II. Even after accounting for the fact that the LOBO test I accuracies on target classes might be affected by chance (since it was not repeated 100 times) it is interesting to see that both tests have almost the exact same accuracy on target bot classes.

\subsection{Improvements with small data additions}

There is a silver lining that is clearly noticeable in this evaluation.
Apparently, it does not matter how much data of a bot class is added in proportion to the dataset size, improvement in performance follows.

However, there is one more important fact. In the LOBO test II, we are testing the classifier on the complete target class. This means, potentially, that adding 500 bots to a simple classifier allows us to further detect the full botnet (in this case, 357,000 instances) at over 99\% accuracy up from 62\%. If we further delve into the details, we can see from the learning speed test that adding 16 samples gets us to 99\% accuracy on the Star Wars bots. 

\subsection{Scalability}

While re-training a classifier several times can be computationally expensive, we have empirically shown that using a few examples on a dataset with balanced bot classes yields similar and stable results. Reducing the size of each bot class from several thousands to 500 decreases the needed resources significantly.
Resource-wise, this specific implementation of the classifiers used a relatively large amount of storage capacity, mostly because we are analysing \emph{all} the tweets for each of the users. This ends up being several terabytes of data. However, most, if not all, of the methodology can be implemented in parallel. Collecting, parsing, sampling, training (each) classifier, and testing can easily be done in parallel.

The importance of this method goes beyond this, as it can readily allow multiple bot classes to be plugged in as needed, provided there are more than a few samples of them. 

\section{Conclusion}

In this paper, we investigated the resilience of bot detection systems on Twitter. 
We showed that these systems perform very well when trained on homogeneous data, but that their performance drops dramatically when they are tested on classes of bots that they have not observed before.
We also proposed a methodology to evaluate how well we can expect any given classifier to generalize on unseen bot data.
It uses different bot datasets or classes as a proxy for the new and unseen classes.
These unseen classes may be developed in the future, but may also be already present but undetected.
This finding has important implications for our research field, since it shows that detection systems might not generalize very well, a problem that becomes particularly important in the fast paced and inherently adversarial world of social network abuse.

\descr{Acknowledgments.} This project has received funding from the European Union's Horizon 2020 Research and Innovation program under the Marie Sk\l{}odowska-Curie ENCASE project (Grant Agreement No. 691025).

\small
\bibliographystyle{abbrv}
%\bibliography{refs}

\end{document}